\documentclass[9pt,twocolumn,twoside]{osajnl}
\journal{ol} % Choose journal (ao, aop, josaa, josab, ol, pr)

\setboolean{shortarticle}{true}
\usepackage[justification=justified]{caption}

\title{Temporal modulation of a spectral compressor for efficient quantum storage}

\author[1]{Karthik V. Myilswamy}
\author[1,*]{Andrew M. Weiner}

\affil[1]{School of Electrical and Computer Engineering and Purdue Quantum Science and Engineering Institute, Purdue University, West Lafayette, Indiana 47907, USA}
\affil[*]{Corresponding author: amw@purdue.edu}

\begin{abstract}
Spectral and temporal mode matching are required for the efficient interaction of photons and quantum memories. In our previous work \cite{myilswamy2020spectral}, we proposed a new route to spectrally compress broadband photons to achieve spectral mode matching with narrowband memories, using a linear, time-variant optical cavity based on rapid switching of input coupling. In this work, we extend our approach to attain temporal mode matching as well by exploiting the time-variation of output coupling of the cavity. We numerically analyze the mode matching and loss performance of our time-varying cavity and present a possible implementation in integrated photonics.
\end{abstract}

\setboolean{displaycopyright}{true}
\begin{document}
\maketitle

%Quantum networks are of prime importance in distributed quantum information processing and long-distance quantum communication \cite{kimble2008quantum}. 
Quantum networks generally involve different physical platforms like matter-based systems for implementing %quantum 
memory nodes and photons for long-distance transportation  \cite{kimble2008quantum}. Hence, we must address the challenge of interfacing these platforms. 
%The efficiency of interconnection between these platforms is contingent on the spectral and temporal mode matching between them. 
In addition to the quantum  frequency conversion \cite{Tanzilli2005, rutz2017quantum} required to match the center wavelengths, we need to achieve spectral and temporal mode matching which is the focus of this work. The memories are generally narrowband with linewidths on the order of a few hundred MHz or below. Whereas the photons generated from typical spontaneous parametric down-conversion sources are broadband with linewidths on the order of a few hundred GHz or more. The temporal shape of the incoming photon also plays a role in the probability of it being absorbed by the quantum memory. Hence, spectral compression along with temporal engineering is necessary to efficiently interface broadband photons with narrowband memories. %nodes. 

Prior approaches to spectral compression include nonlinear optical schemes \cite{lavoie2013spectral,allgaier2017highly,li2017spectral} and time-lens operations \cite{karpinski2017bandwidth, sosnicki2018large, sossnicki2020large}. The conventional approaches to temporal shaping are electro-optic intensity modulation and nonlocal shaping of heralded photons \cite{baek2008temporal,averchenko2020efficient} and are lossy even in an ideal scenario. In our previous work \cite{myilswamy2020spectral}, we proposed a novel approach of using time-varying cavities to achieve spectral compression. Our approach is based on capturing a broadband photon into a cavity via rapid switching of the input coupling. By using high-$Q$ cavities, our approach enables the possibility of compressing photons to a sub-GHz bandwidth, limited only by the cavity linewidth. This allows spectral mode matching between broadband photons and narrowband memories. However, the temporal mode of the compressed photons is a pulse train with a decaying exponential envelope, which generally is not optimal for matching to a quantum memory. Hence in this work, we propose to employ the time-variation of output coupling of the same cavity to simultaneously engineer the temporal shape of the spectrally compressed photons. Our approach is theoretically lossless.

Quantum memory demonstrations with on-demand storage and retrieval are generally three-level systems and fall under the category of $\Lambda$-type schemes \cite{guo2019high, wei2020broadband, saglamyurek2021storing,wang2019efficient}.
%\cite{wang2019efficient,wei2020broadband, saglamyurek2021storing, Michelberger2015, guo2019high, cho2016highly}. %A number of different phenomena like electromagnetically induced transparency \cite{wang2019efficient, wei2020broadband}, Autler-Townes splitting \cite{wei2020broadband, saglamyurek2021storing}, off-resonant Raman transition \cite{Michelberger2015, guo2019high} and gradient echo memory \cite{cho2016highly} have been explored in these demonstrations. 
In these quantum memories, an auxiliary classical control pulse is used to mediate the on-demand storage and retrieval of quantum signals. One can either optimize the temporal shape of the classical control pulse or the quantum signal relative to the other for maximizing the interaction efficiency \cite{gorshkov2008photon, novikova2008optimal, guo2019high}. References \cite{guo2019high, wei2020broadband, saglamyurek2021storing} demonstrate efficient quantum storage of quantum
signals with near-Gaussian temporal modes with widths on the order of few ns to few tens of ns. Deriving motivation from these references, we explore and analyze the possibility of achieving such Gaussian-shaped temporal modes for the spectrally compressed photons  for the initial investigation of this idea. However, our analysis is extendable to any generic temporal mode of interest.%, depending on the system in-hand. % Of these memory demonstrations, particular interests to us are references \cite{guo2019high, wei2020broadband, saglamyurek2021storing}, in which they have demonstrated efficient quantum storage of broadband quantum signals with near-Gaussian temporal modes with widths in the order of few ns to few tens of ns. Hence in this work, we explore and analyze the possibility of achieving a Gaussian temporal mode for the spectrally compressed photons that are output from our time-varying cavity.%Finally, we discuss a potential integrated photonics implementation that can produce Gaussian-shaped photons of a few ns width, compatible with relatively broadband quantum memories.

\begin{figure}
    \centering
    \includegraphics[trim=0 340 365 0,clip,width=3in]{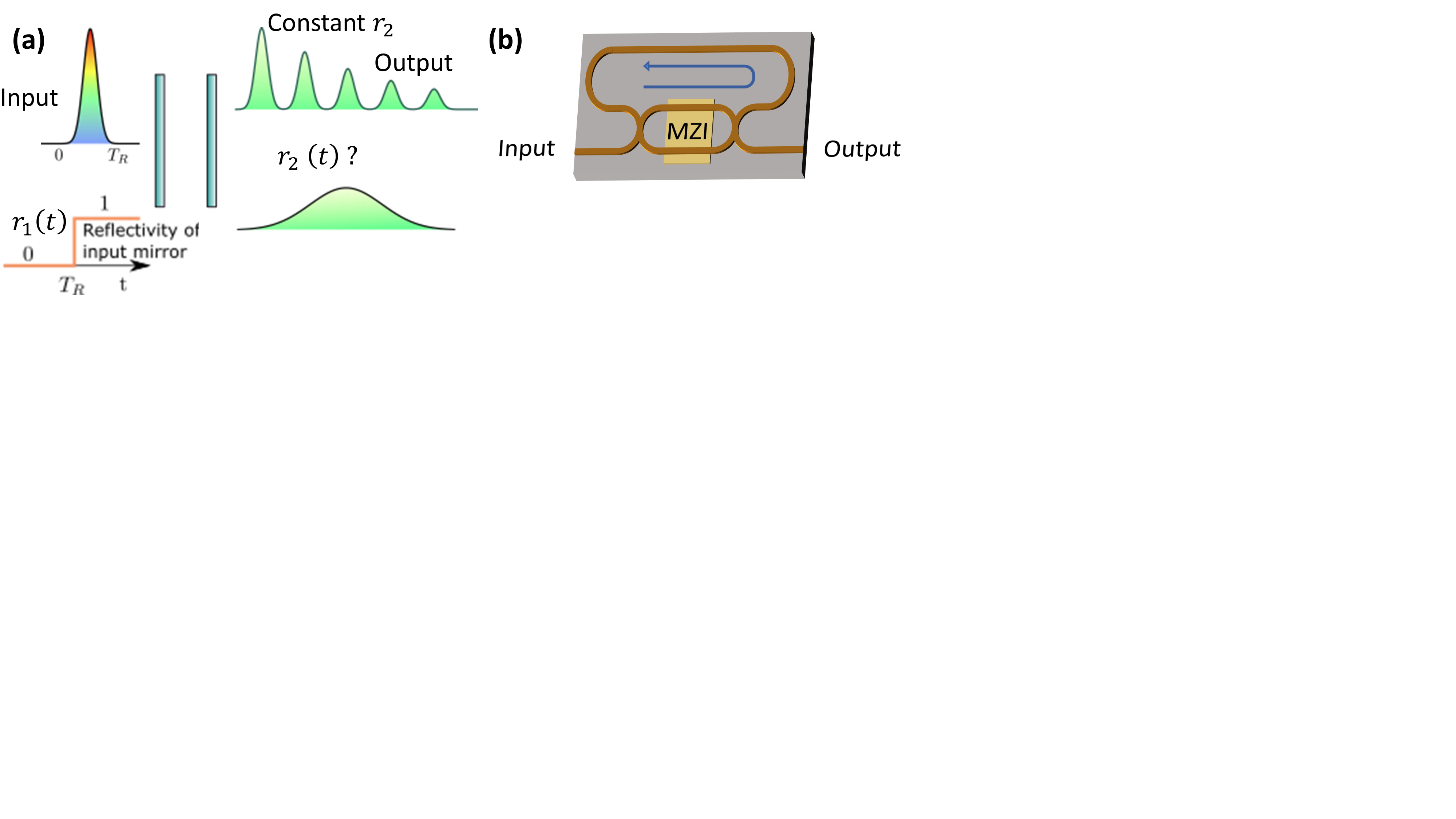}
\caption{(a) Time-varying cavity with input reflectivity rapidly switched from 0 to 1 after the pulse enters the cavity. For a constant output reflectivity $r_2$, it leads to a spectrally compressed pulse train with a decaying exponential temporal envelope. The time-variation of $r_2(t)$ can be used to reshape the output temporal mode. (b) Integrated photonics implementation.}
\label{fig1}
\end{figure}

We consider a Fabry-Perot (FP) cavity [Fig.~\ref{fig1}(a)] for our discussions, although our concept is applicable to any generic cavity structure. We also
show a possible integrated photonics implementation using a microring configured with a Mach-Zehnder interferometer (MZI) in Fig.~\ref{fig1}(b), discussed later. Consider a FP cavity with the input and output field reflection (transmission) coefficients given by $r_1 (t_1)$ and $r_2 (t_2)$, respectively. $r_1$ is rapidly switched from $0$ to $1$ at $t=T_R$, where $T_R$ represents the roundtrip time. For input pulses contained between $t=0$ and $t=T_R$, the entire pulse energy is captured inside the cavity and exits only through the output mirror. In the case of a constant $r_2$, the output is a pulse train containing time-shifted copies of the input with a decaying exponential envelope. If the input carrier frequency matches with one of the cavity modes, most of the energy gets compressed into that resonance linewidth. The rapid switching of $r_1(t)$ ensures energy confinement resulting in this spectral compression. With the additional usage of time-variation of $r_2(t)$, we now seek to engineer the temporal mode of the spectrally compressed output. In principle, one may seek to use a suitable $r_2(t)$ to attain any arbitrary temporal mode. % However, for the initial investigation of this idea, here we confine our analysis to Gaussian temporal modes.  
As long as the time variation of $r_2(t)$ is sufficiently slow compared to the roundtrip time, we intuitively expect the spectral compression properties to be largely preserved, consistent with our simulation results presented later.
%In theory, one must be able to use suitable $r_2(t)$ to attain any arbitrary temporal mode of interest; the amount of spectral compression is proportionally related to the temporal width. However, we confine our analysis to Gaussian temporal shape for the investigation of this idea.

\begin{figure}
    \centering
    \includegraphics[trim=0 150 0 0,clip,width=3.5in]{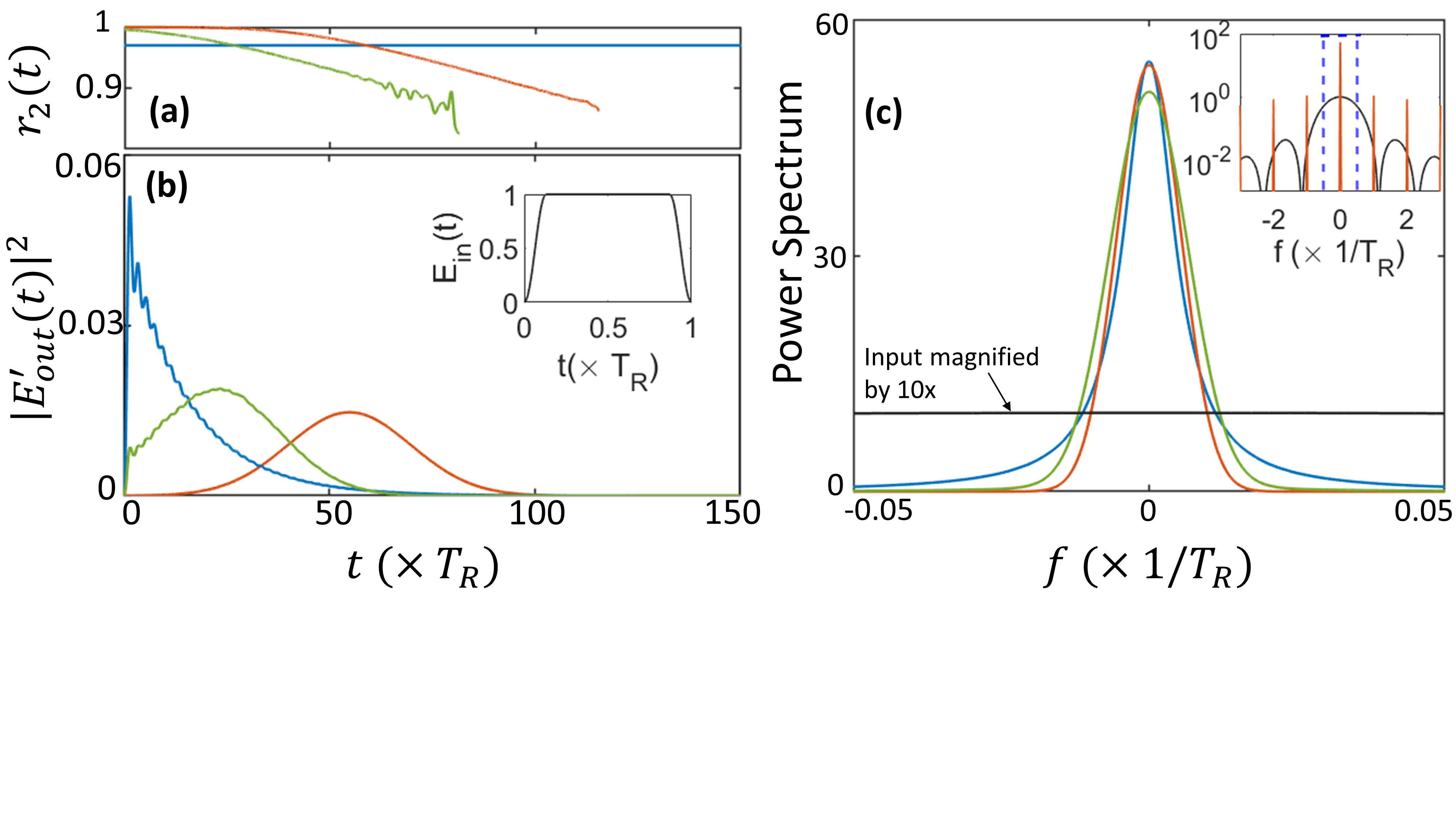}
\caption{For the three different cases of $r_2(t)$ shown in (a), (b) temporal and (c) spectral profiles of the output $E_{out}^{\prime}$ are plotted. The input pulse is shown in the inset of (b) and $r_1(t)$ is assumed to be abruptly switched from 0 to 1 at $t=T_R$. $2\alpha L=0.025$ dB is assumed. The blue curve corresponds to a constant $r_2(t) = 0.97$ and the other two colors are the optimal solutions for specific conditions (details in text). In (c), the input is shown in black, and plots are normalized to maximum input power spectral density. FWHM of the input is $\approx 1/T_R$. In the inset of (c), the log plots of the power spectrum are shown for both input (not magnified) and $E_{out}$ (corresponding to red curve) to highlight the multiple peaks that are filtered to obtain $E_{out}^{\prime}$. Dashed lines represent the filtering operation.}
\label{fig2}
\end{figure}

We use $E$ for representing the oscillating electric fields. The output field $E_{out}(t)$ of such a cavity for any generic input $E_{in}(t)$ can be expressed using an iterative relation \cite{crosignani1986time,sacher2008dynamics} given by:
\begin{equation}
\label{eq1}
    \begin{aligned}
        \frac{E_{out}(t)}{t_2(t)} & = e^{-\alpha L/2}~~ t_1\left( t- \frac{T_R}{2}\right) ~ E_{in} \left( t -\frac{T_R}{2} \right) \\[0.3ex]
       & + e^{-\alpha L}~ ~r_1\left(t-\frac{T_R}{2}\right) ~r_2\left(t-T_R\right) ~\frac{E_{out}\left(t-T_R \right)}{t_2\left(t-T_R\right)} \\[1ex]
       E_{out}(t) &  = 
        \Bigg\{ \sum_{m=0}^{\infty} e^{-\alpha L (1/2 + m)}~ t_1\left[t-\left(mT_R + \frac{T_R}{2} \right) \right]t_2(t)  \\[0.3ex]
       & E_{in}\left[t-\left(mT_R + \frac{T_R}{2} \right) \right]\prod_{n=0}^m b_n \Bigg\} \\
      b_0 &= 1, \;\;\;  b_n = r_1\left[t-\left(nT_R - \frac{T_R}{2} \right) \right]r_2\left(t-nT_R\right) \;\;\forall n \geq 1 
    \end{aligned}
\end{equation}
where $L$ represents the roundtrip length and $2\alpha L$ the roundtrip loss. For any given $E_{in}(t)$, $r_1(t)$ (rapid switching from 0 to 1), and $\alpha L$, one can try to find a $r_2(t)$ such that the output $E_{out}(t)$ resembles the targeted Gaussian shape $E_{tar}(t) = e^{-t^2/\sigma^2}e^{j\omega_0 t}$, where $\omega_0$ is the carrier frequency of the input and $\sigma^2$ determines the temporal width. Assuming excessive loss is to be avoided, the achievable temporal widths are upper bounded by the intrinsic cavity lifetime.
Hereon we assume $E_{in}(t)$ oscillates at one of the cavity resonances, into which the spectral compression occurs. % $\omega_0$ is set to 0 for plotting purposes without loss of generality. 
We define two parameters, namely, fidelity ($\zeta$) and efficiency ($\eta$) as figures of merit to quantify the system performance. Fidelity captures the similarity between the produced output profile and the targeted output profile, given by the following integral.
\begin{equation}
\label{eq2}
    \zeta = \frac{\left|\int E_{out}^{'}(t) E_{tar}^{*}(t) dt \right|^2}{\int \left|E_{out}^{'}(t)\right|^2 dt \int \left|E_{tar}(t)\right|^2 dt  }
\end{equation}
$E_{out}^{'}(t)$ is obtained by filtering out the spectral content of $E_{out}(t)$ that falls outside of the cavity resonance which coincides with the carrier frequency. We use a flat-top bandpass filter of width FSR ($1/T_R$) centered at the corresponding resonance for this filtering operation [shown in inset of Fig.~\ref{fig2}(c)]. This is a valid assumption as we don't expect the narrowband memory to respond to the filtered out spectral content. Efficiency ($\eta$) describes the loss performance of our system or, equivalently, the ratio of energy present in the spectral compression peak to the input:
\begin{equation}
\label{eq3}
    \eta = \frac{\int\left|E_{out}^{'}(t)\right|^2 dt}{ \int \left|E_{in}\right|^2 dt  }
\end{equation}
Both $\zeta$ and $\eta$ take values between 0 and 1, with the ideal scenario being $\zeta=\eta=1$. $\eta$ is upper bounded by 1 as it is a linear and passive system, and $\zeta$ is upper bounded by 1 due to the Cauchy-Schwarz inequality. We use a numerical optimization technique called particle swarm optimization (PSO) \cite{bonyadi2017particle} to obtain an optimal $r_2(t)$ that satisfies Eq.\ref{eq1}, while minimizing the cost function $\mathbb{C}(\zeta,\eta) = \eta\times log_{10}(1-\zeta)$. This cost function provides a higher weight to $\zeta$ than to $\eta$, which allows us to emphasize the temporal shaping ability in our analysis. PSO is easy to implement and does not require the problem to be differentiable.
For optimization simulations, we discretize $r_2(t)$ at integer multiples of $T_R$ starting from $T_R/2$ away from the input time (the instant at which the output starts to take nonzero values); the discretized optimal $r_2(t)$ is later smoothened using spline interpolation for potential practical implementation.

%To illustrate these discussions, we consider a flat-topped input pulse of width $T_p = T_R$ except for its sinusoidal edges accounting for a net 25$\%$ of the pulse width [inset of Fig.~\ref{fig2}(b)]. Here, we refer to the total time over which the input pulse is non-zero as its pulse width $T_p$. 
For our simulations we assume an input pulse time-limited to the pulse duration $T_p$; the field is flat-topped over the central $0.75T_p$, with symmetric sinusoidal rising and falling edges accounting for the remaining $0.25T_p$ [inset of Fig.~\ref{fig2}(b)].  Initially we assume the pulse duration is matched to the roundtrip time: $T_p = T_R$. $E_{in}(t)$ is input between $t=0$ and $t=T_R$, and $r_1(t)$ is switched abruptly from 0 to 1 at $t=T_R$. $2\alpha L = 0.025~$dB is assumed. We have shown both temporal and spectral profiles of the output ($E_{out}^{\prime}$) in Fig.~\ref{fig2}(b) and Fig.~\ref{fig2}(c), respectively, for the three different cases of $r_2(t)$ shown in Fig.~\ref{fig2}(a). For a constant $r_2(t)=0.97$ (blue), the output has a decaying exponential envelope. For the other two cases, we perform numerical optimization to obtain $r_2(t)$ to mode match $E_{out}^{\prime}(t)$ with the targeted Gaussian temporal mode of width $\sigma=30\times T_R$, while minimizing $\mathbb{C}(\zeta,\eta)$. Here and in the following figures, we show $r_2(t)$ only for the values of $t$, over which the output is non-negligible. The red curve corresponds to the optimal case, with $\zeta = 0.9999$ and $\eta = 0.66$. %($-1.8$ dB)
The green curve corresponds to the case when we force the optimizer to not maximize $\zeta$ beyond $0.95$, resulting in a larger $\eta=0.78$.
This reduced fidelity can be connected to the truncated portion of the Gaussian tail near $t=0$. The optimal outputs are essentially of the form $e^{-(t-\tau)^2/\sigma^2}$ with $\tau$ chosen by the optimizer, determining the amount of truncation at the leading edge and consequently the fidelity and the loss. Hence to achieve a high fidelity with the Gaussian shape, most of the input power has to execute a higher number of roundtrips resulting in a higher loss for a cavity with nonzero loss. These two examples demonstrate the trade-off between fidelity and loss performance. Also, the decaying exponential output resulting from the constant $r_2(t)=0.97$ has a comparably low $\zeta = 0.77$, but has a higher $\eta = 0.82$. %($-0.86$ dB).
All these cases have sharp peaks in their spectral profiles [Fig.~\ref{fig2}(c)] at the center indicating spectral compression (spectral narrowing along with spectral peak enhancement). The output spectral widths are similar for the three cases and are roughly two orders lesser than the input. This shows the ability of our system to both spectrally compress and temporally shape the incoming photon.%, achieved by the time-variation of input and output coupling of the cavity, respectively.
\begin{figure}
    \centering
    \includegraphics[trim=10 135 90 0,clip,width=3.2in]{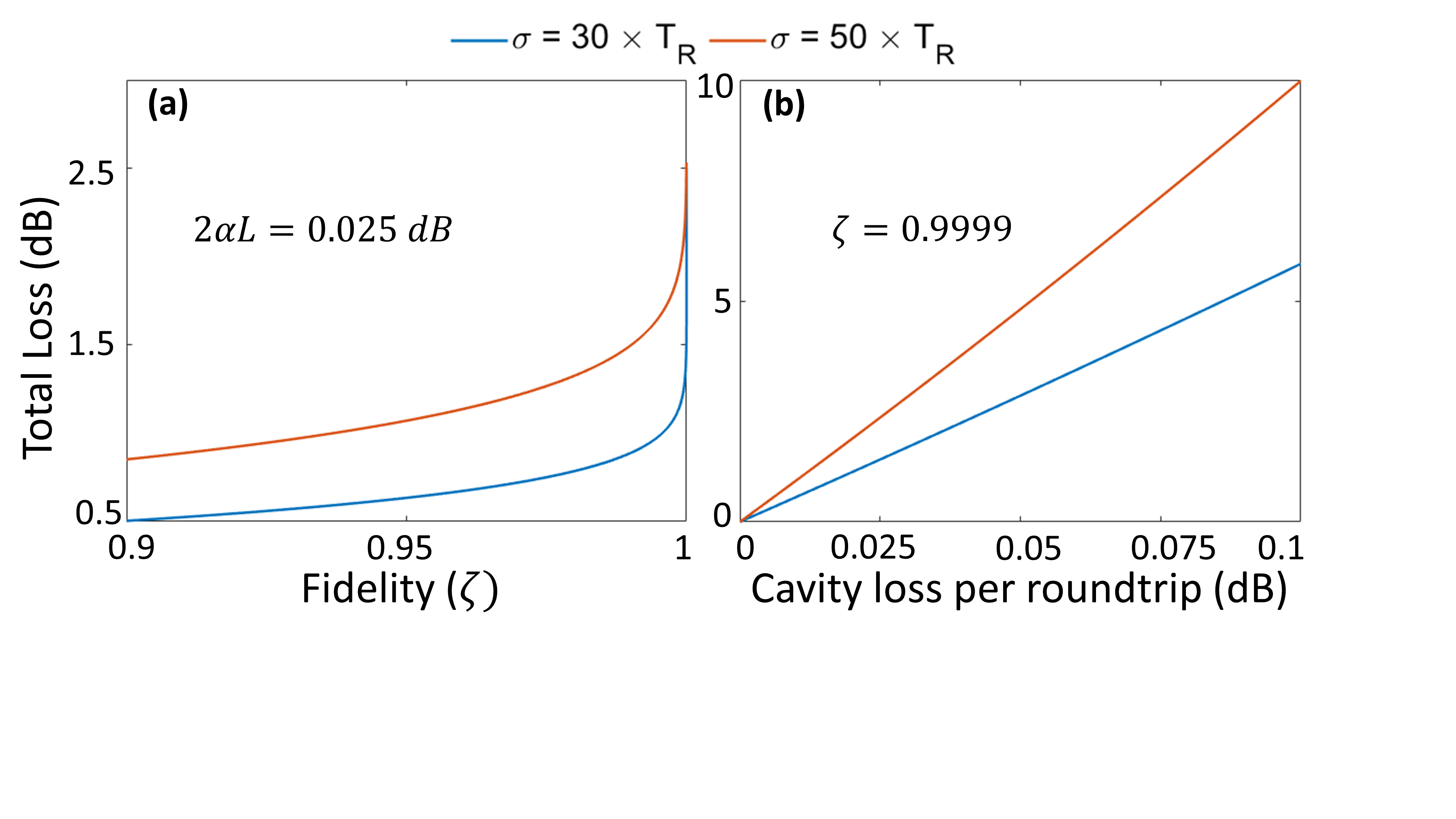}
\caption{Estimated loss resulting due to cavity attenuation for different values of (a) fidelity (assuming $2\alpha L = 0.025$~dB) and (b) cavity roundtrip loss (while maintaining $\zeta=0.9999$) to achieve the Gaussian temporal shapes of widths $\sigma \in \{30\times T_R, 50 \times T_R\}$.}
\label{fig3}
\end{figure}

To further expand on the trade-off between fidelity and loss, we consider  Gaussian profiles of the form $e^{-(t-\tau)^2/\sigma^2}$ over $t\geq 0$. If $\tau$ is intentionally varied, the outputs are similar except with different amounts of the leading edge truncated at $t=0$, which leads to different values of the fidelity $\zeta$ [see red and green curves in Fig.~\ref{fig2}(b)]. We can analytically estimate the loss due to cavity attenuation, even without specifically writing down the required $r_2(t)$ function using the formula: $\eta = \sum_{n=0}^{\infty} |E(t_n)|^2/e^{\alpha L} \sum_{n=0}^{\infty} |E(t_n)|^2 e^{2\alpha L t_n/T_R}$. Here the output field is represented by samples corresponding to n roundtrips through the cavity, each of which has experienced n+1/2 roundtrips worth of cavity attenuation.
However, this estimate does not include the other sources of loss (shorter input pulses, $T_p<T_R$, finite risetime for $r_1$), which are considered later. In Fig.~\ref{fig3}(a), we plot the estimated total loss ($-10log_{10}\eta$) resulting solely due to attenuation in the cavity (cavity loss) for different values of $\zeta$ to achieve the Gaussian temporal modes of widths $\sigma \in \{ 30 \times T_R, 50 \times T_R \}$, for $2\alpha L = 0.025$ dB. The total loss increases with the increase in fidelity, consistent with the examples given above, illustrating the trade-off. Also, the total loss is higher for achieving a Gaussian of larger temporal width as it requires most of the input power to execute an increased number of roundtrips. This loss estimate can be seen as a lower bound to achieve certain fidelity values for any input as it excludes the other practical loss sources mentioned above. In Fig.~\ref{fig3}(b), we plot the estimated total loss to achieve $\zeta = 0.9999$ as a function of roundtrip attenuation. The total loss increases with the increase in roundtrip attenuation and is higher for achieving the output modes with larger temporal width, as expected. One can choose an optimal $\tau$ based on this analysis and can reverse engineer the required $r_2(t)$, constraining to the physical laws. However, it is simpler to use numerical optimization to arrive at the solution in an efficient manner.
%Given this analysis, one can operate our time-varying cavity as required for their applications at a specific fidelity or loss performance.
\begin{figure}[!ht]
    \centering
    \includegraphics[trim=5 655 30 0,clip,width=3.2in]{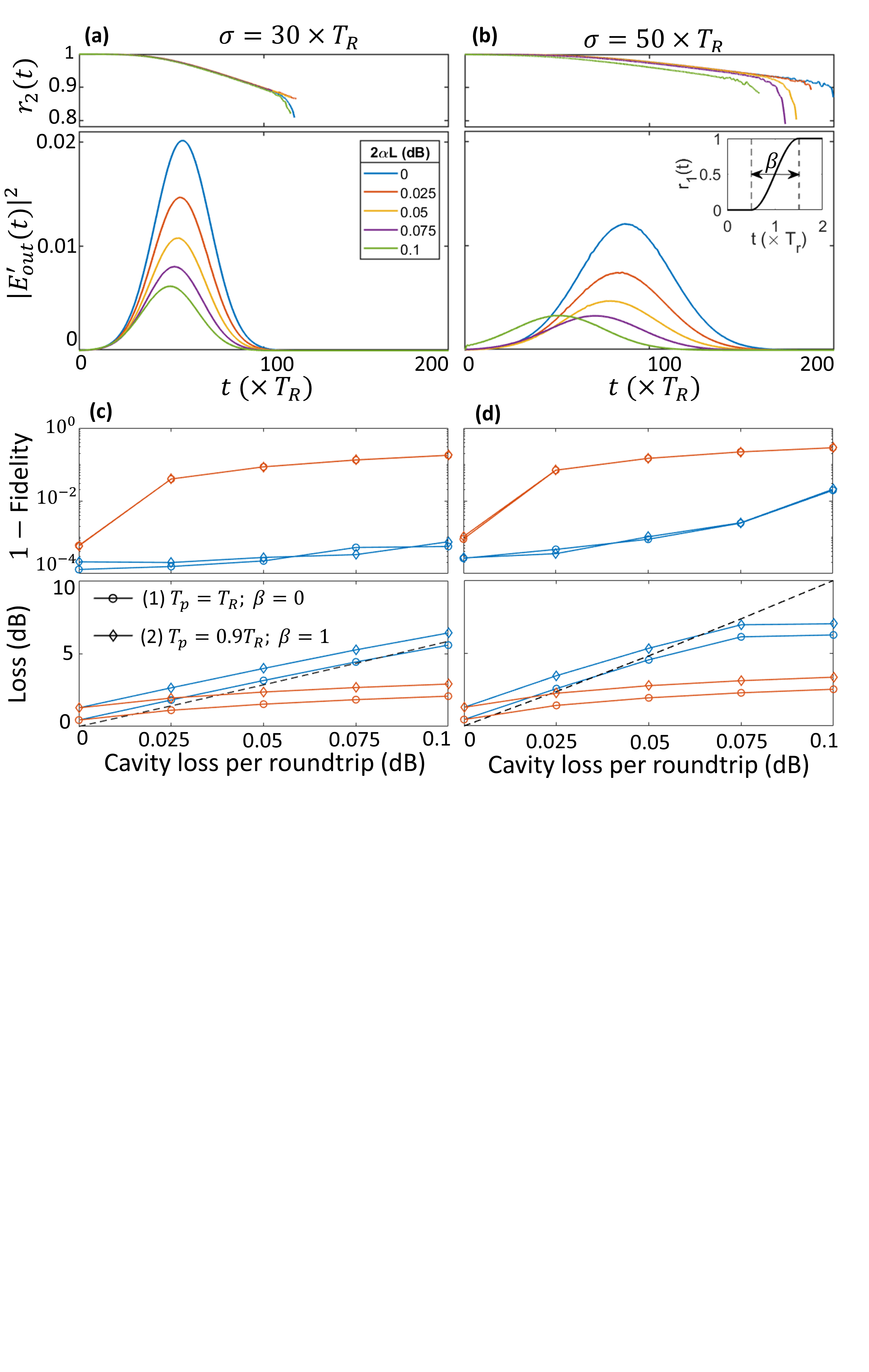}
    \caption{Optimization results for achieving Gaussian temporal modes of widths $\sigma = 30 \times T_R$ (left column) and $\sigma = 50\times T_R$ (right column). $r_1(t)$ is assumed to be a raised cosine function as shown in inset of 4(b). We consider two cases: (1) $T_p = T_R, \beta = 0$ and (2) $T_p = 0.9T_R, \beta = 1$. For case (1), the optimal $r_2(t)$ minimizing $\mathbb{C}(\zeta,\eta)$ and the corresponding output profiles are shown in (a,b) for different values of cavity loss. For both cases, the fidelity and loss metrics are plotted in (c,d) for the cost functions $\mathbb{C}(\zeta,\eta)$ (blue) and $\mathbb{C}^{\prime}(\zeta,\eta)$ (red). Dashed lines in (c,d) correspond to the estimated loss for $\zeta=0.9999$ plotted in Fig.~\ref{fig3}(b) for respective temporal widths.}
\label{fig4}
\end{figure}

We now analyze the performance addressing practical scenarios for different values of cavity loss, rise time for switching $r_1$, and input pulse width. $r_1(t)$ is assumed to be a raised cosine function [inset of Fig.~\ref{fig4}(b)], switching from 0 to 1 with the transition centered at $t=T_R$ over the rise time $\beta T_R$. The input pulse is as defined earlier, but with $T_p$ not necessarily equal to $T_R$.  We consider two different cases: (1) $T_p = T_R, \beta = 0$ and (2) $T_p = 0.9T_R, \beta = 1$. For input pulses shorter than $T_R$, substantial power gets compressed into the other cavity modes, effectively leading to additional loss as only the central mode is assumed to overlap with the memory. In the case of a nonzero rise time for switching $r_1$, some power is lost to the input side. For a nonzero $\beta$, the input arrival time needs to be optimized additionally to account for power lost to the input side. Here the optimizer is again allowed to choose the optimum value of the output pulse delay ($\tau$). For these cases, we discuss the performance metrics to achieve the targeted Gaussian temporal modes of widths $\sigma = 30 \times T_R$ [Fig.~\ref{fig4} left column] and $\sigma = 50 \times T_R$ [Fig.\ref{fig4} right column]. For case (1), the optimal $r_2(t)$ minimizing $\mathbb{C}(\zeta,\eta)$ and the corresponding $\left|E_{out}^{\prime} (t)\right|^2$ for achieving the Gaussian temporal shapes of widths $\sigma = 30 \times T_R$ and $\sigma=50 \times T_R$ are shown in Fig.~\ref{fig4}(a) and Fig.~\ref{fig4}(b), respectively, for different values of roundtrip loss. For both cases (1) and (2), the fidelity and loss ($-10log_{10}\eta$) are shown in Figs.~\ref{fig4}(c,d) as a function of roundtrip loss (blue curves). The fidelity drops and the loss increases with the increase in roundtrip loss. As the cavity loss increases, the optimizer settles for Gaussian profiles with increased truncation at the leading edge to minimize the loss by reducing the total number of roundtrips, resulting in the fidelity drop. The fidelity performance is similar for both cases. Although the loss is higher in case (2) compared to case (1) as power leaked into the adjacent resonances and the input side loss increase due to $T_p = 0.9T_R$ and $\beta = 1$, respectively, the difference is only $\sim1$ dB. 

We have considered a cost function with a higher weight to $\zeta$ so far to emphasize the shaping ability. However, a cost function of the form $\mathbb{C}^{\prime}(\zeta,\eta)=1-\zeta \eta$, with equal weights to $\zeta$ and $\eta$, is directly related to the probability of the photon being captured by the quantum memory, as %which is directly proportional to the product of $\zeta$ and $\eta$.
$\zeta$ describes the overlap integral between the photon and the memory and $\eta$ represents the total loss (similar to Ref.~\cite{novikova2008optimal}). We have plotted the fidelity and loss for this cost function in red in Figs.~\ref{fig4}(c,d) for the cases analyzed above. This leads to lower fidelity but also lower loss compared to the former cost function. In Figs.~\ref{fig4}(c,d), the dashed line corresponds to the estimated loss for achieving $\zeta=0.9999$ from Fig.~\ref{fig3}(b). It can be observed that the optimizer settled to a loss lower than this estimate by sacrificing on $\zeta$ especially for $\mathbb{C}^{\prime}(\zeta,\eta)$ and in a few cases of $\mathbb{C}(\zeta,\eta)$ when operated at higher roundtrip loss. If we compare this cavity with optimal time-varying $r_2(t)$ to a cavity with optimal constant $r_2$, this has 3 times higher $\mathbb{C}^{\prime}(\zeta,\eta)$ for Gaussian temporal output modes when the roundtrip loss is zero. However, this contrast in performance decreases with the increase in roundtrip loss. %This discussion illustrates how different parameters affect the performance metrics and serves as a guide to operate our system optimally.

As discussed in our earlier work \cite{myilswamy2020spectral}, one can envision an integrated photonics demonstration using a microring configured with a switchable coupler, formed from a MZI [Fig.~\ref{fig1}(b)]. Initially the MZI is in the cross-state, allowing the input pulse to enter the cavity. However, before the light arrives back at the upper input port of the MZI, the control voltage to the MZI is rapidly switched to a state close to the bar state (effectively high $r_2$).  This captures the input pulse inside the cavity with high efficiency.  Subsequently a small portion of the trapped light is output after every roundtrip, controlled via small variations of the voltage to the MZI.  Such weak out-coupling is analogous to the weak transmission through the partially reflecting output mirror of a FP cavity.  Thus, the time variation of both input and output coupling can be realized using the same MZI.  Alternatively, one may envision a ring resonator with separate input and output (drop) ports, each of which is configured with its own MZI for time-dependent coupling.
%As discussed in our earlier work \cite{myilswamy2020spectral}, one can envision an experimental demonstration in an integrated photonics platform using a ring resonator cavity configured with a switchable coupler, formed from a MZI [Fig.~\ref{fig1}(b)]. The input pulse can be captured inside the cavity with high efficiency by operating the MZI in cross state, after which a small portion of the trapped light is output  after every roundtrip by controlling the MZI accordingly. The coupling of the small fraction of light out of the cavity is analogous to the partially reflecting output mirror of the FP cavity. The time variation of both input and output coupling can be realized using the same MZI.
With recent demonstrations of high-Q cavities, and high-speed electro-optic modulators, thin-film lithium niobate (TFLN) is a promising platform to realize this demonstration  \cite{Zhu2021}. The record-low losses reported in TFLN platforms are $\sim2.7~$dB/m for straight waveguide sections \cite{wang2018integrated}, and $\sim0.15~$dB/cm for phase-shifter sections with a $V_{\pi}\cdot L$ value of $2.325$~V-cm and a $3~$dB bandwidth of $\geq67~$GHz \cite{Xu2020}.
As an example we consider a $6.5~$mm long TFLN cavity, which corresponds to a 50 ps roundtrip time - longer than available switching times but much shorter than achievable photon lifetimes.  If the phase shifter section is 2.5 mm long, then the roundtrip loss should be $\sim0.05$ dB.
%If we consider a TFLN cavity of length $6.5~$mm, out of which $2.5~$mm accounts for the phase shifter, this corresponds to a roundtrip time of $\sim50~$ps and a roundtrip loss of $\sim0.05~$dB. The corresponding choice can lead to spectrally compressed pulses with widths in the order of a few ns, which would be of interest for efficient quantum storage.
%A 3~dB bandwidth of $67~$GHz \cite{Xu2020} % or $175~$GHz \cite{kharel2021breaking}
%corresponds to a rise time of a few ps, enabling rapid switching compared to the roundtrip time of $50~$ps. 
The dispersion in LN resonators can be considered negligible for the input pulses whose widths are around the roundtrip time \cite{he2018dispersion}. For the input pulse shown in the inset of Fig.~\ref{fig2}(b) of width $50~$ps, it would be possible to achieve the Gaussian temporal modes of FWHM ($\sqrt{2ln2}~\sigma$) of $1.8~$ns and $3~$ns with losses $3~$dB and $4.5~$dB, respectively, with $\zeta\approx 1$.
%with $\zeta\approx 1$ and $-10log_{10}\eta = 3~$dB, and $3~$ns with $\zeta\approx 1$ and $-10log_{10}\eta = 4.5~$dB.
If the phase shifter loss can be reduced to the level of the straight waveguide section, the roundtrip loss significantly drops to $\sim0.018~$dB. Consequently, it will be possible to achieve a lower loss of $1.4~$dB, $2~$dB, and $2.9~$dB for the Gaussian temporal modes of FWHM $1.8~$ns, $3~$ns, and $6~$ns, respectively, with $\zeta\approx 1$.

In summary, we have proposed a way to incorporate temporal engineering capability in our novel electro-optic approach to spectral compression using time-varying cavities. Our system is linear and hence is applicable for both classical and quantum light. Providing for a bandwidth interface of broadband photons to narrowband quantum memories will contribute to new opportunities for entanglement distribution, such as entanglement swapping between ground stations linked by a satellite.

%\smallskip
\textbf{Funding.} National Science Foundation (ECCS-1809784); Air Force Office of Scientific Research (FA9550-20-1-0283)

\textbf{Acknowledgements.} The authors thank Hsuan-Hao Lu for introducing us to PSO and insightful discussions.

\textbf{Disclosures.} The authors declare no conflicts of interest.
\bibliography{References}
\bibliographyfullrefs{References}
\end{document}